\documentclass[aps,prd,twocolumn,nofootinbib]{revtex4-1}
\usepackage{epsfig}
\usepackage[colorlinks,linkcolor=blue,anchorcolor=blue,citecolor=blue,urlcolor=blue,breaklinks=true]{hyperref}
\usepackage{dcolumn}
\usepackage[all]{hypcap}
\usepackage{graphics}
\usepackage{color}
\usepackage{amsmath}
\usepackage{float}
\usepackage{bm}
\usepackage[justification=raggedright]{caption}
\usepackage{subcaption}
\begin{document}
\title{Chiral phase transition in  QED$_3$ at finite temperature}
\author{Wei Wei$^{1}$, Hai-Xiao Xiao$^{1}$, and Hong-Shi Zong$^{1,2,3}$}\email[]{Email: zonghs@nju.edu.cn}
\address{$^{1}$Department of Physics, Nanjing University, Nanjing 210093, China\\
$^{2}$Joint Center for Particle, Nuclear Physics and Cosmology, Nanjing 210093, China\\
$^{3}$State Key Laboratory of Theoretical Physics, Institute of Theoretical Physics, CAS, Beijing, 100190, China}
\begin{abstract}
Chiral phase transition in (2+1)-dimensional quantum electrodynamics (QED$_3$) at finite temperature is investigated in the framework of truncated Dyson-Schwinger equations (DSEs). We go beyond the widely used instantaneous approximation and adopt a method that retains the full frequency dependence of the fermion self-energy. We also take further step to include the effects of wave-function renormalizations and introduce a minimal dressing of the bare vertex. Finally, with the more complete solutions of the truncated DSEs, we revisit the study of chiral phase transition in finite-temperature QED$_3$.
\bigskip

\end{abstract}
\maketitle

\section{INTRODUCTION}
QED$_3$ is a theoretical field model that has been extensively investigated~\cite{Phys.Rev.Lett.60.2575, Phys.Rev.Lett.62.3024, PhysRevD.44.540, Phys.Lett.B.253.246, Phys.Lett.B.266.163, Phys.Lett.B.295.313, Nucl.Phys.B.386.614, PhysRevD.50.1068,1126-6708-1999-03-020}. And many insightful results have been obtained in simplified models~\cite{PhysRevD.90.036007,*PhysRevD.90.065005,*PhysRevD.86.105042}. One of the major interests in various studies of QED$_3$ is the well-known dynamical chiral symmetry breaking (DCSB). Previous studies have shown the existence of DCSB in zero-temperature QED$_3$~\cite{PhysRevLett.60.2575}. Similar conclusion has also been confirmed in QED$_3$ at finite temperature~\cite{PhysRevD.50.1068}. Since QED$_3$ possesses properties such as confinement ~\cite{PhysRevD.46.2695,PhysRevD.52.6087} that are similar to those of quantum chromodynamics (QCD), it is generally believed that investigations into the phase structure of finite-temperature QED$_3$ provide insights into its counterpart in QCD, which is important, for example, in the study of the early Universe. On the other hand, QED$_3$ also finds application in condensed matter systems. It has been suggested by Dorey \emph{et al} that the parity-invariant QED$_3$ could serve as an effective long-wave model for certain two dimensional condensed matter system ~\cite{Nucl.Phys.B.386.614}. Furthermore, QED$_3$ also seems to be relevant to the graphene problem in the continuum limit ~\cite{doi:10.1142/S0217979207038022}. The experimental discovery of the Dirac fermions ~\cite{novoselov2005two,*zhang2005experimental} makes QED$_3$ more than a toy model for QCD. Quantitative studies are thus necessary to clarify the relation between the theoretical concepts and the experimental results.

Strictly speaking, there is no spontaneous symmetry breaking (SSB) in (2+1) dimensional quantum systems at finite temperature, due to the infrared singularities associated with the Goldstone sector as mandated by Coleman-Mermin-Wagner theorem ~\cite{coleman1973,PhysRevLett.17.1133} (see also, for example, Appendix \ref{A}). Nevertheless, in condensed matter, the mean field transition temperature provides a correct energy scale below which the order parameter becomes finite and the spatial correlation becomes strong and long-ranged. Moreover, in a realistic layered system, the inter-layer coupling can easily drive the system into a true ordered state once the in-plane correlations are already strong, e.g., below the mean field transition temperature. In particular, for a U(1) or O(2)
symmetry to be broken, there is in fact an algebraic order below the so-called Kosterlitz-Thouless transition temperature, a temperature not far from the mean field one.
Calculations presented in this paper go beyond the mean field approximation and they also suggest the existence of a second-order phase transition.

For the theory with chiral symmetry, the perturbative approach could not introduce couplings between the two helicity components $\psi_L$ and $\psi_R$ of the fermion field and thus a fermion mass. Therefore,  the DCSB is a nonperturbative phenomenon. The continuum approach based on DSEs is capable of describing the dynamics mass generation in QED$_3$ ~\cite{ROBERTS1994477,doi:10.1142/S0218301303001326}. In this paper, we shall adopt a new ansatz based on the truncated DSEs and apply it to the study of chiral phase transition in QED$_3$ at finite temperature.

As mentioned above, applications to condensed matter systems make it even more interesting to investigate QED$_3$ at finite temperature. However, studies of finite-temperature QED$_3$ are further complicated by two problems compared to the zero-temperature case: First, the O(3) symmetry of QED$_3$ at zero temperature is explicitly broken to the O(2) symmetry at finite temperature in the Euclidean space formulation. The temporal integration is then replaced by a summation over Matsubara frequencies, which leads to considerable complexity in computation. The second problem is the infrared divergence caused by the absence of magnetic thermal screening~\cite{Nucl.Phys.B.386.614, Lo2011164}. In many literature, both of the above problems are avoided by adopting the replacement $\Delta_{\mu\nu}(q_0,\boldsymbol{q})\to \Delta_{00}(0,\boldsymbol{q})$ ~\cite{Nucl.Phys.B.386.614, Aitchison199291, doi:10.1142/S0217732307021251, PhysRevD.86.065002}, where $\Delta_{\mu\nu}(q_0,\boldsymbol{q})$ is the full finite-temperature photon propagator. This approximation drops the spatial component of the photon propagator and assumes only instantaneous interaction. However, the temporal photon acquires a finite mass at finite temperature and becomes short-ranged, while the spatial photon remains massless and long-ranged.
It has been shown by Liu \emph{et al} that the mass of photon tends to suppress the dynamical mass generation ~\cite{PhysRevD.67.065010}.
Therefore, there is no reason why the role of spatial photon should be diminished. On the other hand, instantaneous approximation is also problematic as the leading contribution to the infrared behavior of fermion self-energy is dominant only at high temperature, while the dynamic mass generation usually happens at a much lower temperature, rendering the approximation unjustified.

Since approximations neglecting contributions from the interaction with spatial component of photon or assuming only instantaneous interaction could possibly lead to unreliable results, we shall attempt to to relax the above two approximations in our calculation. We also take into consideration the full effects of wave-function renormalizations. Furthermore we go beyond the widely adopted rainbow approximation by introducing a minimal dressing to the bare vertex. The more sophisticated model enables a more reliable study of chiral phase transition in finite-temperature QED$_3$.

\section{Notation and general equations}
In this paper, we adopt the conventions
\begin{align}
p = (p_0,\boldsymbol{p}),\ \ p_0 = \frac{(2l+1)\pi}{\beta},\ \ p^2=p_0^2+\boldsymbol{p}^2,\\
k = (k_0,\boldsymbol{k}),\ \ k_0 = \frac{(2n+1)\pi}{\beta},\ \ k^2=k_0^2+\boldsymbol{k}^2,
\end{align}
where $l$ and $n$ are integers. $p$ and $k$ denote the external and internal momentum of the fermion propagator, respectively, while $q=p-k$  represents the momentum of the photon propagator. All the equations will be formulated in Euclidean space.

The DSEs at finite temperature can be shown to be
\begin{align}\label{eq:dsef}
S^{-1}(p_0,\boldsymbol{p}) =&\ S_0^{-1}(p_0,\boldsymbol{p})\nonumber\\
&+\frac{e^2}{\beta} \sum_{ k_0 = -\infty}^{+\infty}\int \frac{d^2 \boldsymbol{k}}{(2\pi)^2} \gamma_{\sigma}S(k_0,\boldsymbol{k})\Gamma_{\nu}\Delta_{\sigma\nu}
\end{align}
\begin{align}\label{eq:dsep}
\Pi_{\sigma\nu}(q_0,\boldsymbol{q})=-\frac{N_f e^2}{\beta} \sum_{ k_0 = -\infty}^{+\infty}\int \frac{d^2 \boldsymbol{k}}{(2\pi)^2}\gamma_{\sigma}S(k_0,\boldsymbol{k})\Gamma_{\nu}S(p_0,\boldsymbol{p}) 
\end{align}
where $S(p_0,\boldsymbol{p})$ is the full fermion propagator, $\Delta_{\sigma\nu}$ is the full photon propagator and $\Gamma_{\nu}$ is the full vertex, while $S_0(p_0,\boldsymbol{p})$ and $\gamma_{\sigma}$ are the bare fermion propagator and vertex respectively. We will set $\alpha\equiv N_f e^2 / 8 = 1$ in our calculation.

At finite temperature, the O(3) symmetry is explicitly broken and then we are led to the full fermion propagator of the form 
\begin{align}\label{eq:full_fermion}
S^{-1}(p_0,\boldsymbol{p},\beta) = &\ i A(\boldsymbol{p}^2,p_0^2,\beta) \boldsymbol{p}\cdot \boldsymbol{\gamma}
+i C(\boldsymbol{p}^2,p_0^2,\beta) p_0\gamma_0\nonumber\\
& + B(\boldsymbol{p}^2,p_0^2,\beta).
\end{align}

The full photon propagator has the form
\begin{align}\label{eq:propa}
\Delta_{\mu\nu}(q_0,\boldsymbol{q},\beta) = \frac{P^{L}_{\mu\nu}}{q^2+\Pi_A} + \frac{P^{T}_{\mu\nu}}{q^2+\Pi_B} - \xi\frac{ q_{\mu}q_{\nu}}{q^4},
\end{align}
where
\begin{align}
P^{L}_{\mu\nu} =& \left(\delta_{\mu0}-\frac{q_{\mu}q_0}{q^2}\right)\frac{q^2}{\boldsymbol{q}^2}\left(\delta_{0\nu}-\frac{q_0q_{\nu}}{q^2}\right), \\
P^{T}_{\mu\nu}=&\ \delta_{\mu i}\left(\delta_{ij}-\frac{q_iq_j}{\boldsymbol{q}^2}\right)\delta_{j\nu},
\end{align}
are longitudinal and transverse projection operators, respectively, satisfying $P^{L}_{\mu\nu}+P^{T}_{\mu\nu}=\delta_{\mu\nu}-q_{\mu}q_{\nu}/q^2$, while $\Pi_A$ and $\Pi_B$ are the corresponding photon self-energies. We choose to work in Landau gauge $\xi=0$ in this paper.

For vertex we choose the ansatz $\Gamma_{\nu}(p,k)=f(p,k)\gamma_{\nu}$, with $f(p,k)=(A(p)+A(k))/2$. We expect that this ansatz incorporates basic features of a nonperturbative vertex ~\cite{1126-6708-1999-03-020}. In fact , the dressing of the bare vertex is consistent with the corresponding term in Ball-Chiu (BC) vertex at finite temperature ~\cite{strickland1998non} and is analogous to the often used BC1 (first term BC) vertex at zero temperature.

Accurate solution of photon self-energies poses great numerical challenge. In this paper, we approximate the mass of photon in Eq. \eqref{eq:propa} with the perturbative one-loop result ~\cite{Nucl.Phys.B.386.614}, which is a widely used approximation ~\cite{PhysRevD.90.036007,*PhysRevD.90.065005,*PhysRevD.86.105042}.

\section{Chiral phase transition of QED$_3$ at finite temperature}
Studies of chiral phase transition employing DSEs rely heavily on truncation schemes. Application of finite-temperature QED$_3$ to condensed matter systems calls for more accurate description of the chiral phase transition. This section is devoted to determining the location and the order of the chiral phase transition in QED$_3$ at finite temperature based the full solution of Eqs. \eqref{eq:dsef} and \eqref{eq:dsep}.

There are several equivalent choices for order parameter. In this paper, we shall use the fermion chiral condensate defined by
\begin{align}\label{eq:condensate}
\langle \bar{\psi} \psi\rangle =&\ -\frac{1}{\beta}\sum_{k_0=-\infty}^{\infty}\int\frac{d^2\boldsymbol{k}}{(2\pi)^2}\text{Tr}[S(k_0,\boldsymbol{k})]\nonumber\\
=&\ -\frac{1}{\beta}\sum_{k_0=-\infty}^{\infty}\int\frac{d^2\boldsymbol{k}}{(2\pi)^2}\nonumber\\
&\times\frac{B(k_0,\boldsymbol{k})}{C^2(k_0,\boldsymbol{k})k_0^2+A^2(k_0,\boldsymbol{k})\boldsymbol{k}^2+B^2(k_0,\boldsymbol{k})}.
\end{align}

\begin{figure*}[!ht]
  \begin{subfigure}[b]{0.49\textwidth}
    \includegraphics[width=\textwidth]{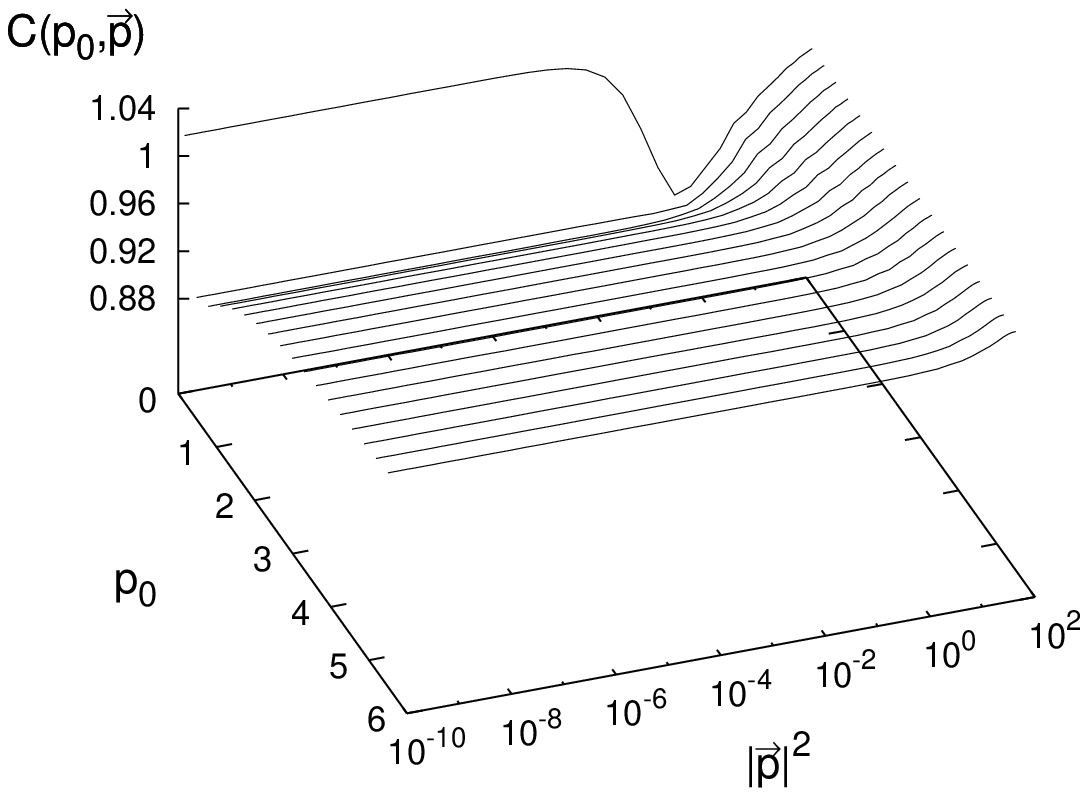}
    \caption{$C(p_0,\boldsymbol{p})$ (T = 0.05)}
  \end{subfigure}
 \hfill
  \begin{subfigure}[b]{0.49\textwidth}
    \includegraphics[width=\textwidth]{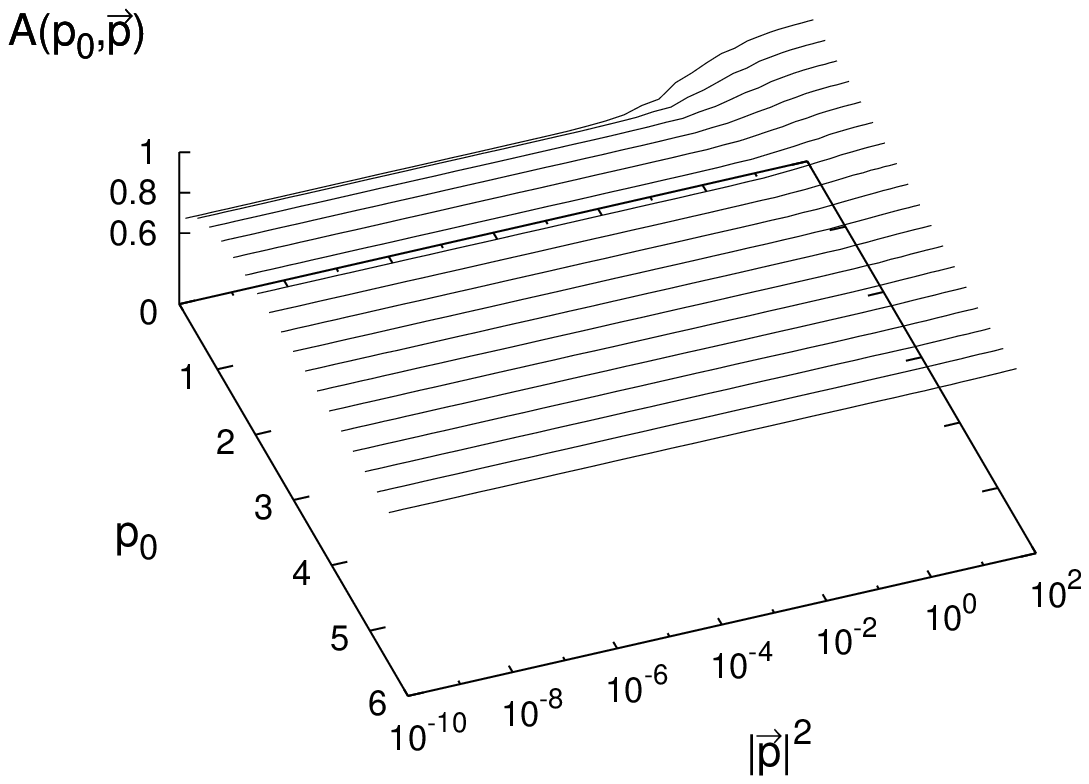}
    \caption{$A(p_0,\boldsymbol{p})$ (T = 0.05)}
  \end{subfigure}
  \begin{subfigure}[b]{0.49\textwidth}
    \includegraphics[width=\textwidth]{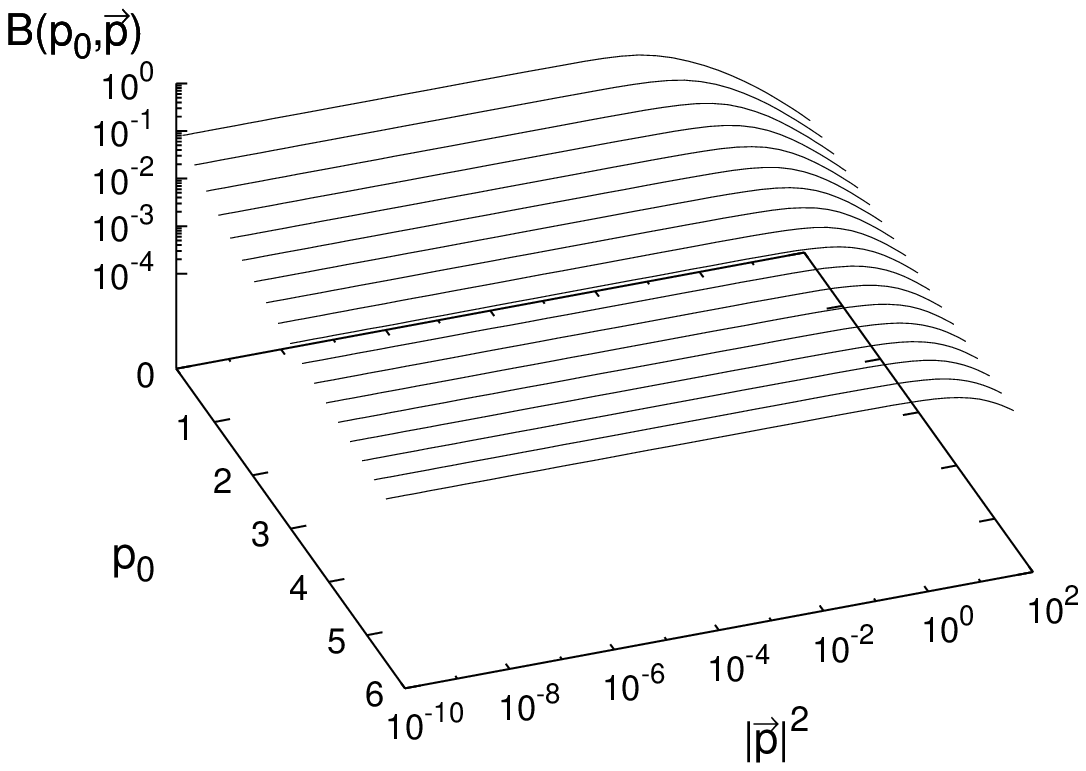}
    \caption{$B(p_0,\boldsymbol{p})$ (T = 0.05)}
  \end{subfigure}
 \hfill
  \begin{subfigure}[b]{0.49\textwidth}
    \includegraphics[width=\textwidth]{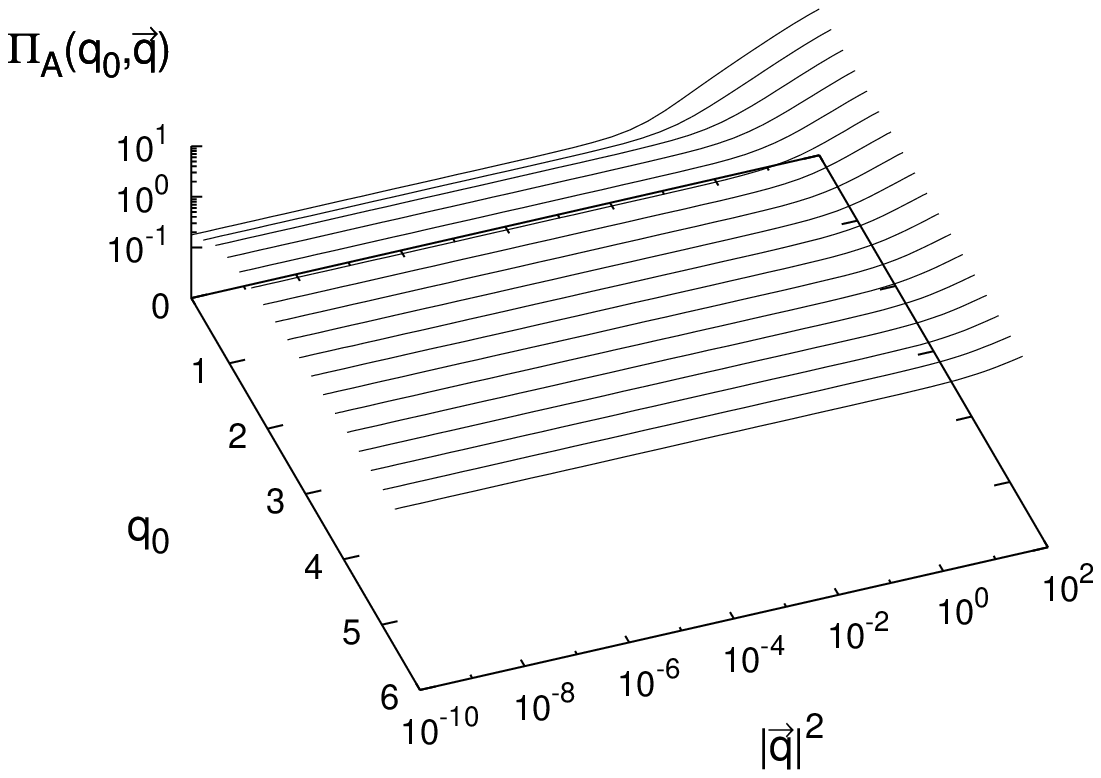}
    \caption{$\Pi_A(q_0,\boldsymbol{q})$ (T = 0.05)}
  \end{subfigure}
  \begin{subfigure}[b]{0.49\textwidth}
    \includegraphics[width=\textwidth]{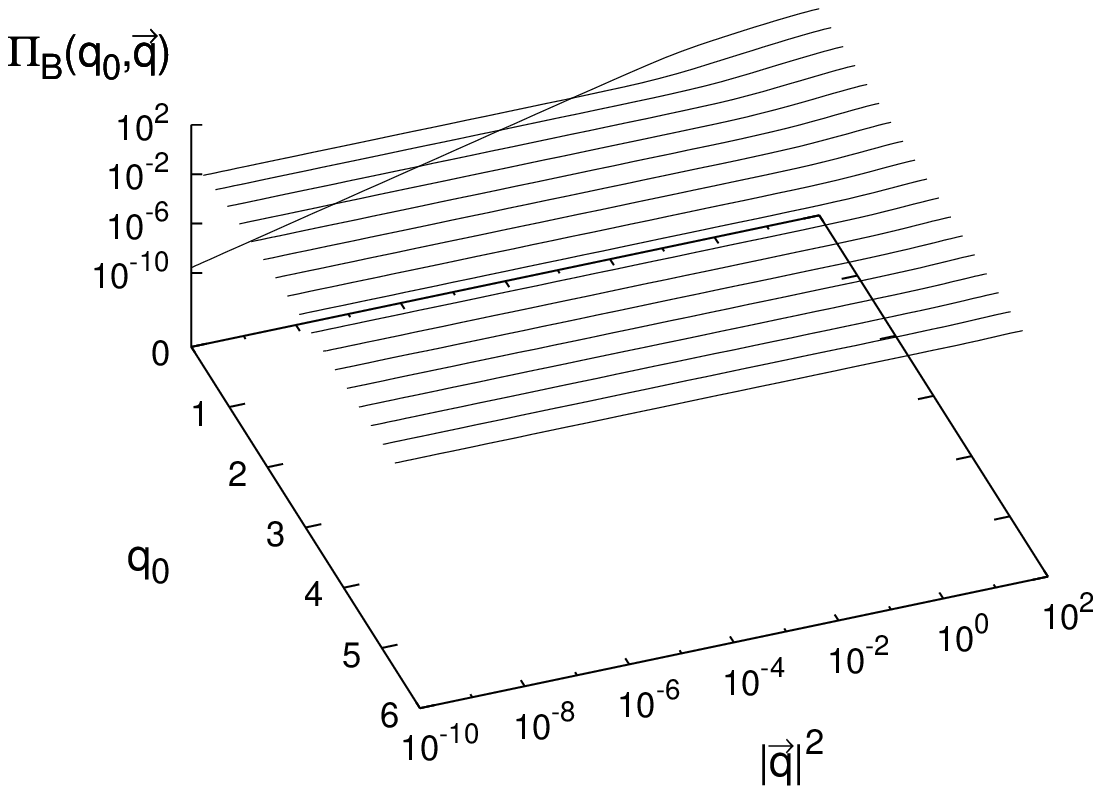}
    \caption{$\Pi_B(q_0,\boldsymbol{q})$ (T = 0.05)}
  \end{subfigure}
\caption{Wave-function renormalizations and self-energies for $N_f = 1$ and $\mu = 0.1$.}
\label{fig:we1}
\end{figure*}

To determine the order of the chiral phase transition, we also need to calculate the thermal susceptibility defined as the response of the fermion chiral condensate to an infinitesimal change of temperature,
\begin{align}\label{eq:susceptibility}
\chi^{T}=\frac{\partial\langle\bar{\psi}\psi\rangle}{\partial T}.
\end{align}

\begin{figure}
\includegraphics[width=\linewidth]{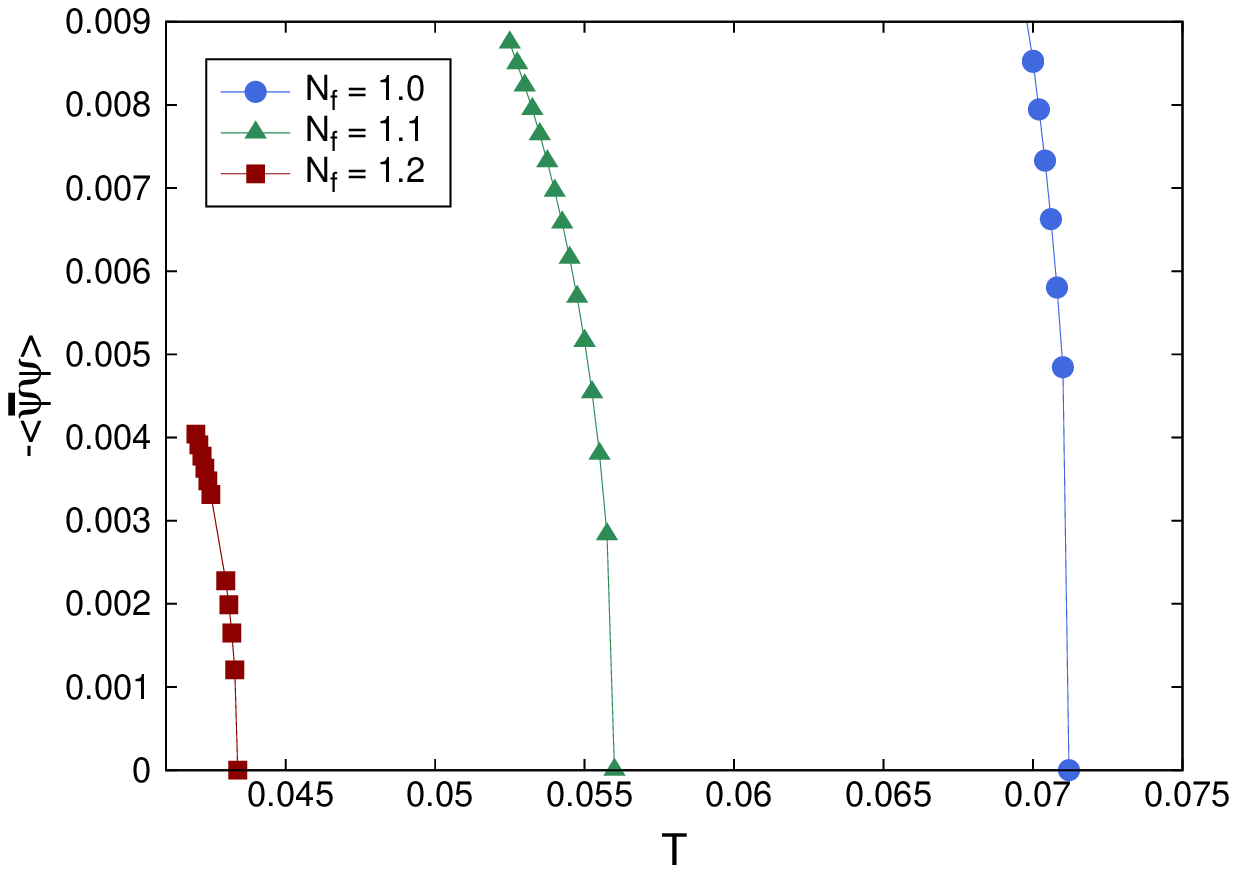}
\includegraphics[width=\linewidth]{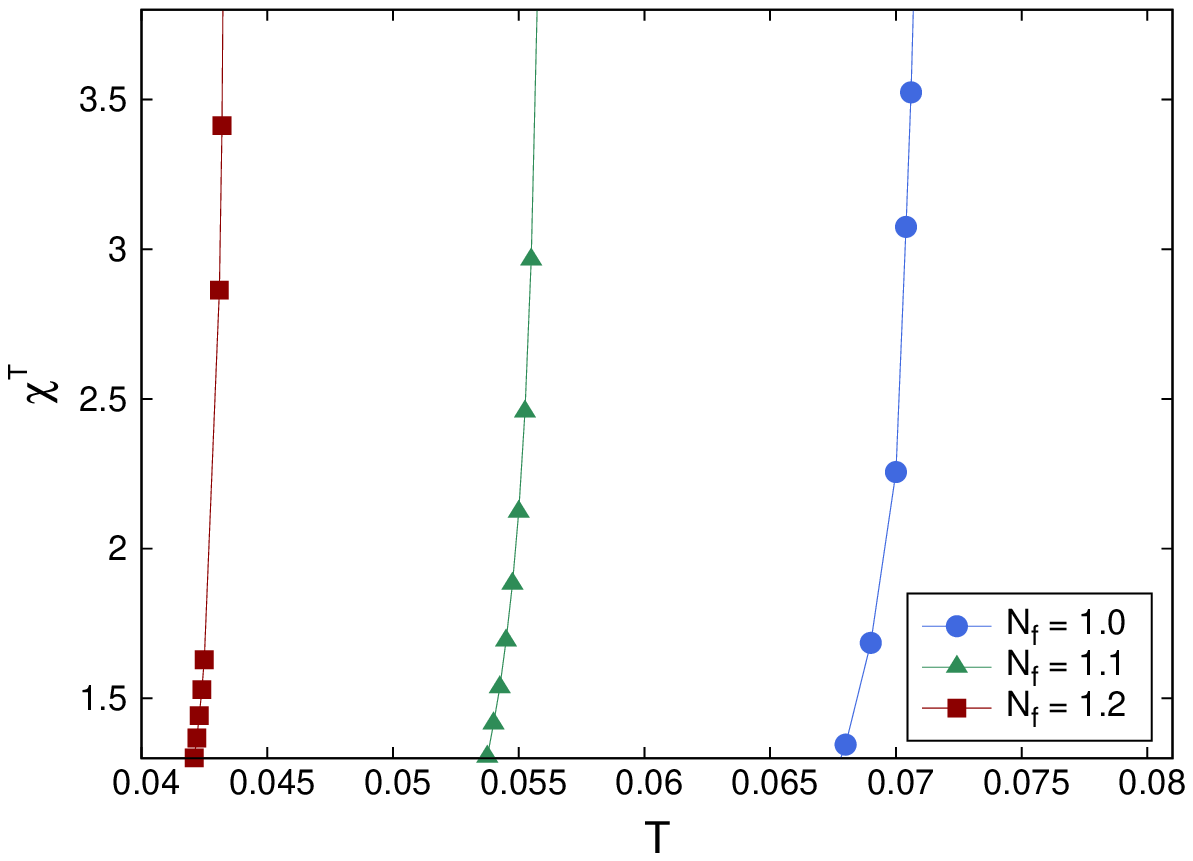}
\caption{Temperature dependence of fermion chiral condensate (\emph{upper panel}) and thermal susceptibility (\emph{lower panel}).}
\label{fig:cds_sus}
\end{figure}

Solving the DSEs for fermion and photon (Eqs. \eqref{eq:dsef} and \eqref{eq:dsep}) numerically, we obtain the wave-function renormalizations and self-energies for different fermion flavor $N_f$. The results are shown in Fig. \ref{fig:we1}. Substituting the results into the the expression for the fermion chiral condensate and the thermal susceptibility, we can then determine the critical behavior of the chiral phase transition. The results are shown in Fig. \ref{fig:cds_sus} and Table \ref{tab:crit}.

The behavior of $\langle \bar{\psi}\psi\rangle$ indicates the existence of chiral symmetry breaking and restoration. Furthermore, the divergence of $\chi^{T}$ near the critical temperature shows that the transition is a second order one. The fermion chiral condensate is suppressed with increasing number of fermion flavor. It is not clear whether there is a critical fermion flavor number $N_c$ beyond which there is no dynamical mass generation at any temperature as it is very hard to examine the low temperature behavior of the theory when the computation is quite involving. And this problem is subjected to future investigation. 

A comparison to one of the mostly used truncations is given in Appendix \ref{B}.

\renewcommand{\arraystretch}{1.5}
\begin{table}
\caption{\label{tab:crit} Dependence of the critical temperature $T_c$ on the fermion flavor number $N_f$.}
\begin{ruledtabular}
\begin{tabular}{cccc}
$N_f$ & 1.0 & 1.1 & 1.2\\
\hline
$T_c$ & $7.12\times10^{-2}$ & $5.58\times10^{-2}$ & $4.34\times10^{-2}$ \\
\end{tabular}
\end{ruledtabular}
\end{table}

\section{summary}
In this paper, we study the critical behavior of the QED$_3$ at finite temperature employing the continuum approach based on DSEs. Due to the Coleman-Mermin-Wagner theorem ~\cite{coleman1973,PhysRevLett.17.1133}, it is generally believed that there is only spontaneously symmetry in QED$_3$ in the zero-temperature limit. Nevertheless, in the large-N$_f$ approximation, the long-range fluctuations are absent and thus dynamical mass generation can still be observed in two dimensions as is illustrated by Gross-Neveu model ~\cite{PhysRevD.10.3235}. Results obtained in the large-N$_f$ approximation provide the correct energy scale below which the order parameter becomes finite,  and from this perspective they are meaningful. Our calculation going beyond the large-N$_f$ approximation and it also suggests the existence of a chirally symmetry broken phase. Comparison to previous conclusions deduced from the large-N$_f$ approximation provides further insights into the finite-temperature QED$_3$. We solve directly the DSEs for fermion and photon propagators with a nonperturbative truncation of the full vertex. We obtain much reduced fermion chiral condensate and critical temperature with increasing number of fermion flavor. The relation between the critical temperature $T_c$ and the critical fermion flavor number $N_c$ is interesting and may be subjected to future research. The ansatz and truncation in this paper relax most approximations in previous literature. We abandon the instantaneous approximation and take into account the full effects of wave-function renormalizations. We also go beyond the widely used bare vertex truncation. Therefore we do not envisage material improvement over current analysis until a gauge-symmetry-preserving truncation scheme for the DSEs is employed.

\acknowledgements
This work is supported by the National Natural Science Foundation of China (under Grants No. 11275097, No. 11475085, and No. 11535005).

\appendix

\section{}\label{A}

Consider the following Green's function for a free scalar boson filed in $D = (2+1)$ at finite temperature:
\begin{align}\label{ir}
\Delta(|x|,\tau;T;m)=&T\sum_{k_0=-\infty}^{\infty}\int\frac{d^2 k}{(2\pi)^2}e^{i\boldsymbol{k}\cdot\boldsymbol{x}+ik_0 \tau}\frac{1}{k^2+k_0^2} \nonumber\\
=&\frac{1}{4\pi}\int_0^{\infty}dk k J_0(k|x_{\perp}|)\frac{1}{\sqrt{k^2+m^2}}\nonumber\\
&\times\coth\left(\frac{\sqrt{k^2+m^2}}{2T}\right).
\end{align}

For $|x_{\perp}|T\gg 1$, the far infrared domain,
\begin{align}
\coth\left(\frac{\sqrt{k^2+m^2}}{2T}\right)=\frac{2T}{\sqrt{k^2+m^2}}+\frac{1}{3}\frac{\sqrt{k^2+m^2}}{2T}+ ...
\end{align}

Thus the integral in \eqref{ir} will diverge for $T>0$ and $m=0$. Therefore, there are no massless Goldstone bosons in $D=(2+1)$ for $T>0$.

\section{}\label{B}

Replacing the full photon propagator $\Delta_{\mu\nu}(q_0,\boldsymbol{q})$ with $\Delta_{00}(0,\boldsymbol{q})$  and setting $C(p_0,\boldsymbol{p})= A(p_0,\boldsymbol{p})= 1$, we arrive at one of the frequently used truncations~\cite{Nucl.Phys.B.386.614, Aitchison199291, doi:10.1142/S0217732307021251, PhysRevD.86.065002},
\begin{align}
B(p_0,\boldsymbol{p}) =\ & m + \frac{8\alpha T}{N_f}\sum_{n=-\infty}^{+\infty}\int\frac{d^2\boldsymbol{k}}{(2\pi)^2}\frac{B(k_0,\boldsymbol{k})}{k^2+B^2(k_0,\boldsymbol{k})}\nonumber\\
&\times\frac{1}{\boldsymbol{q}^2+\Pi_A(q)},
\end{align}
where $m$ is the current fermion mass.
The above equation shows that $B(p_0,\boldsymbol{p})$ automatically losses dependence on frequency as $p_0$ does not appear on the right-hand side of the equation. We can further simplify the equation by using the identity
\begin{align}
\sum_{n=-\infty}^{\infty}\frac{1}{k_0^2+x^2}=&\sum_{n=-\infty}^{\infty}\frac{1}{[(2n+1)\pi T]^2+x^2}\nonumber\\
&=\frac{\tanh(\frac{x}{2T})}{2xT}.
\end{align}
The summation over the Matsubara frequencies can then be performed analytically to give
\begin{align}\label{eq:sigma_zero}
B(\boldsymbol{p},T)=&\ m + \int\frac{d^2\boldsymbol{k}}{(2\pi)^2}\frac{B(\boldsymbol{k},T)\tanh\frac{\sqrt{\boldsymbol{k}^2+B^2(\boldsymbol{k},T)}}{2T}}{2\sqrt{\boldsymbol{k}^2+B^2(\boldsymbol{k},T)}}\nonumber\\
&\times\frac{1}{\boldsymbol{q}^2+\Pi_A}.
\end{align}

\begin{figure*}
\begin{subfigure}{0.49\textwidth}
    \includegraphics[width=\textwidth]{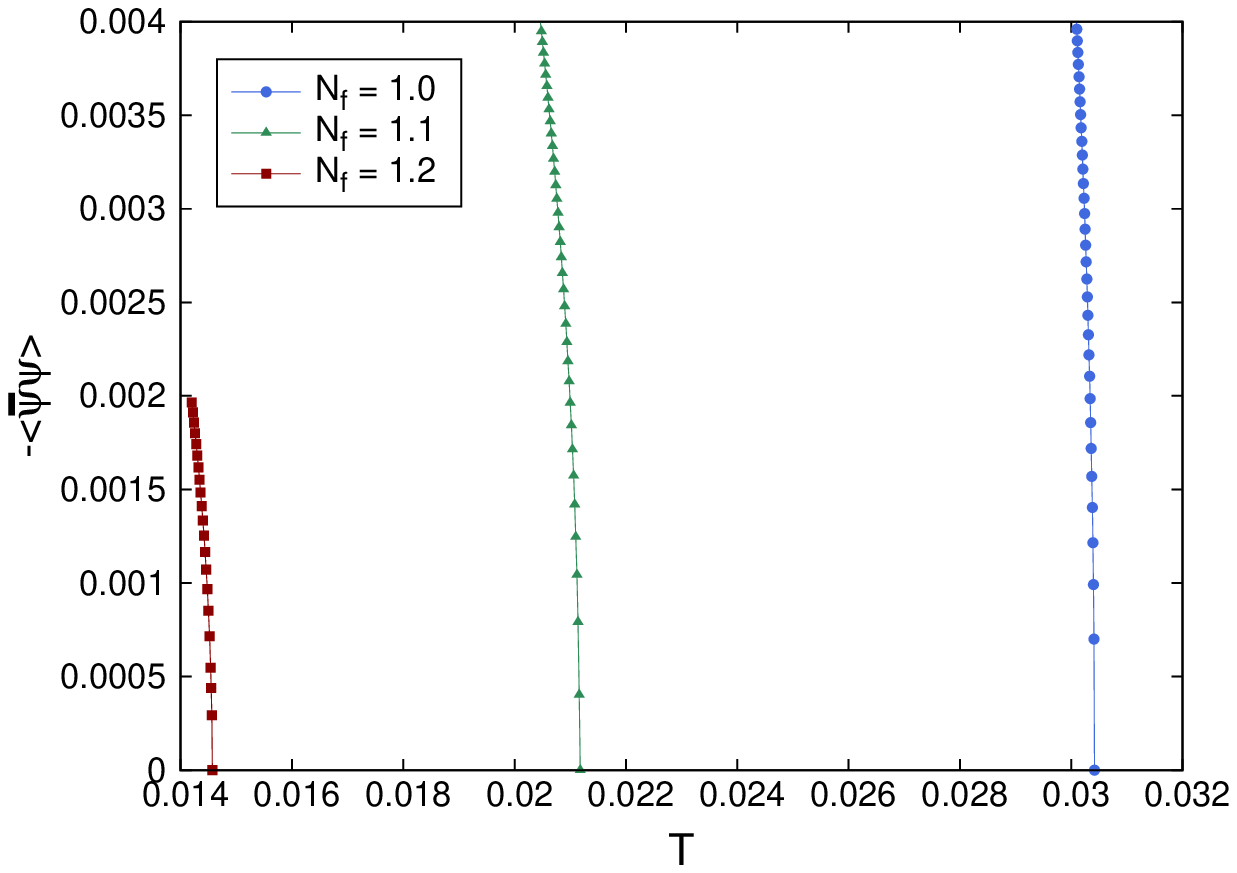}
    \caption{fermion chiral condensate}
  \end{subfigure}
 \hfill
  \begin{subfigure}{0.49\textwidth}
    \includegraphics[width=\textwidth]{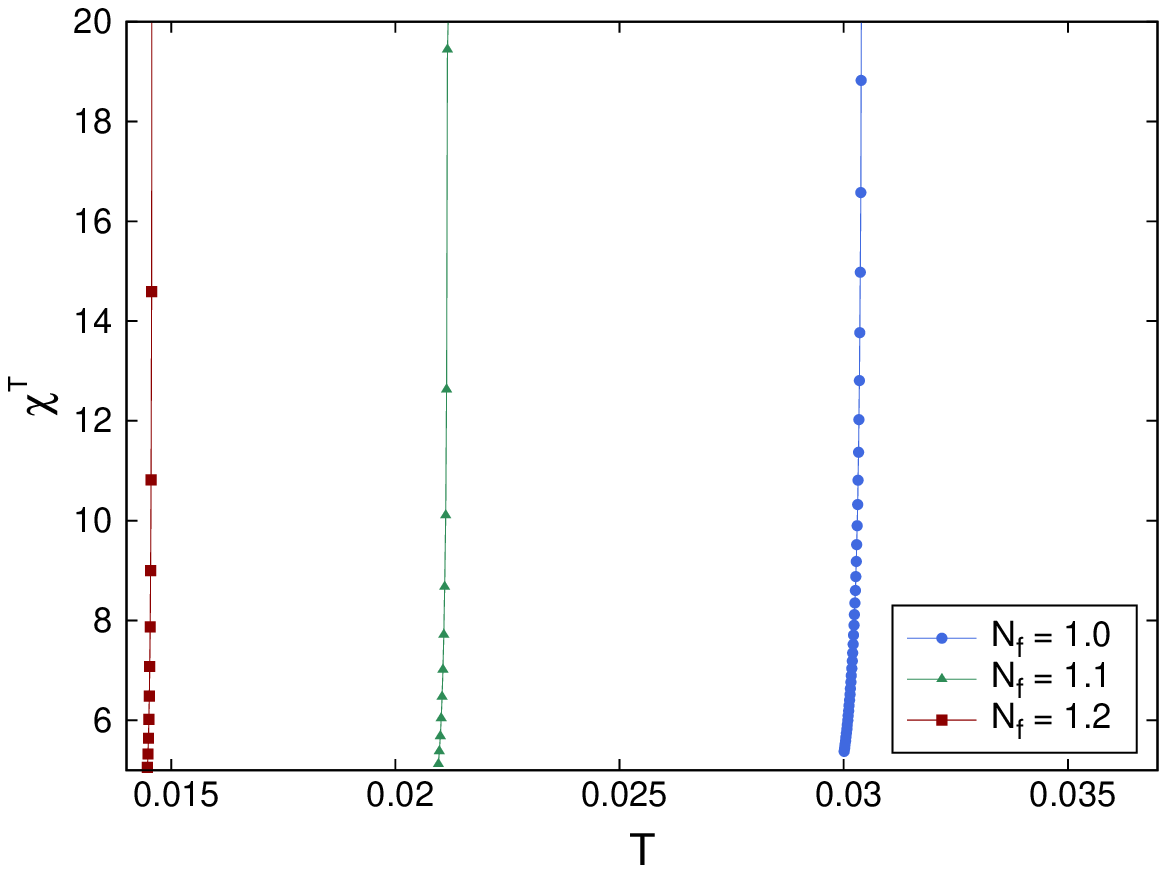}
    \caption{thermal susceptibility}
  \end{subfigure}
\caption{Temperature dependence of fermion chiral condensate and  thermal susceptibility.}
\label{fig:cds_sus_0}
\end{figure*}

With the solution of Eq. \eqref{eq:sigma_zero} in the chiral limit, we can continue to calculate the fermion chiral condensate and thermal susceptibility. The results are shown in Fig. \ref{fig:cds_sus_0} and Table \ref{tab:crit0}.

\renewcommand{\arraystretch}{1.5}
\begin{table}
\caption{\label{tab:crit0} Dependence of the critical temperature $T_c$ on the fermion flavor number $N_f$.}
\begin{ruledtabular}
\begin{tabular}{cccc}
$N_f$ & 1.0 & 1.1 & 1.2\\
\hline
$T_c$ & $3.04\times10^{-2}$ & $2.12\times10^{-2}$ & $1.46\times10^{-2}$ \\
\end{tabular}
\end{ruledtabular}
\end{table}

Comparing Figs. \ref{fig:cds_sus}, \ref{fig:cds_sus_0} and Tables \ref{tab:crit} , \ref{tab:crit0}, we see that the two truncations of DSEs show similar qualitative properties (both calculations suggest a second-order chiral phase transition). However the locations of the chiral phase transition determined by the two truncations are quite different, which suggests that the truncation presented in the appendix is not credible for practical applications (although useful for qualitative studies).

\bibliographystyle{apsrev4-1}
\bibliography{qed3}
\end{document}